\documentstyle[epsfig]{article}
\newcounter{eqnletter}[equation]
\catcode`\@=11
\@addtoreset{equation}{section}   
\catcode`\@=12

\newcommand{\parno}{\par\noindent}

\begin{document}
\begin{center}

{\LARGE \bf  Non-Hermitian tridiagonal random matrices and returns to
  the origin of a  random walk . }
\vskip 1cm
{\large {\bf G.M. Cicuta , M.Contedini} }
\vskip 0.1 cm
Dipartimento di Fisica, Universita' di Parma,\\
and INFN, Gruppo di Parma collegato alla Sezione di Milano\\
Viale delle Scienze,  43100 Parma, Italy \footnote{E-mail
addresses: cicuta@fis.unipr.it \, , \, contedini@fis.unipr.it} \\
\vskip 0.1 cm
\vspace{1 cm}
{\large {\bf L.Molinari}}
\vskip 0.1 cm
Dipartimento di Fisica, Universita' di Milano,\\
and INFN, Sezione di Milano\\
Via Celoria 16 , 20133 Milano, Italy \footnote{E-mail
address : luca.molinari@mi.infn.it }

\vspace{2 cm}

{\large {\bf Abstract }}
\end{center}

\vspace{1 cm}

We study a class of tridiagonal matrix models, the "q-roots of unity"
models, which includes the sign ($q=2$) and the clock ($q=\infty$)
models by Feinberg and Zee. We find that the eigenvalue densities are
bounded by and have the symmetries of the regular polygon with $2 q$ sides,
in the complex plane. Furthermore the averaged traces of $M^k$ are integers that
count closed random walks on the line, such that each site is visited a
number of times multiple of $q$. We obtain an explicit evaluation for
them.

\newpage

\section{Introduction}
Random matrix ensembles are extensively studied since the early works
of Wigner and Dyson, as effective models for the description of statistical
properties of  spectra of complex physical systems, which include 
resonances of heavy nuclei, quantum billiards, mesoscopic transport  and
quenched QCD
 \cite{mehta} -  \cite{msri}. \\

Tridiagonal matrices with random entries naturally occur in the simplest
models of disordered one-dimensional crystals, beginning with the works
by Dyson \cite{dyson} and Schmidt \cite{schmidt}. If the spring constants
are fixed while the masses are random, one is led to  a tridiagonal
Hermitian matrix with random entries on the main diagonal, often referred 
to as a random site problem. \\
More recently, several authors studied the  density
of eigenvalues of non-Hermitian tridiagonal matrices still having the
random entries  in the main diagonal, after the model
introduced by Hatano and Nelson \cite{hatano} to describe the motion of
vortices pinned by columnar defects in a superconductor  .
  In this model, the  spectral density
 was obtained only in the case of a Cauchy probability distribution
for the independent random entries  \cite{zee} , 
 \cite{mudry} : in the large-$N$ limit,
the eigenvalues lie on a "squeezed ellipse"  in the complex plane
with the addition of two "wings" on the real axis, which appear provided
the strength of the random site entries exceeds a critical value
. It was known
for a long time \cite{ginibre} \cite{sommers} that complex non-Hermitian
random matrices may have eigenvalues filling a two-dimensional area
in the complex plane. 
For a  class of tridiagonal non-Hermitian models, with random hopping and
random sites, Goldsheid and Khoruzhenko  \cite{gold} found conditions
for the probability distributions, such that in the large-$N$ limit the eigenvalues
converge to a curve in the complex plane.
Yet, in more general settings it seems still impossible to predict
whether a non-Hermitian ensemble of random matrices has eigenvalues
converging to
 a curve or filling a two-dimensional area  or a fractal.
These difficulties suggest the usefulness of investigating
different types of tridiagonal  non-Hermitian random matrices, possibly by
a variety of techniques, even for models not directly related to a physics
problem.  Feinberg and Zee \cite{fz1} considered a class
of non-Hermitian random hopping models, 
with eigenvalue equation
\begin{eqnarray}
t_{j-1} \, \psi_{j-1}\,+\, s_j^* \, \psi_{j+1} \,= \,E \, \psi_j
\label{i.1}
\end{eqnarray}
and given probability distributions  for the hopping amplitudes  $t_j$ , $s_j$.
In their first model, called the "clock model" , the  hopping amplitudes
are independent random phases.
 They found , numerically, that the
eigenvalues are distributed in a disk in the complex plane, centered at the 
origin, with rotational invariance. In a second model , called the "sign
model" , hopping amplitudes are independent and randomly equal to $\pm1$.
In the case of open chain, it is easy to see that the eigenvalues  are functions of
the $N-1$ products $ t_j \, s_j^*$ , therefore , for the open chain, the
 distribution of eigenvalues of the "sign model" is also obtained by
the study of eigenvalues of the matrix $M(x , 1)$ , given below in eq.(\ref{a.1}),
where the $N-1$ independent  variables $x_i$ are independent  and randomly
 equal to $\pm1$.\\
 In this paper we study a more general model , 
the "q-roots of unity" model, where each independent variable $x_i$ 
is one of the $q$ roots of unity, with uniform probability. Of course,
for $q=2$ , it is again the "sign model". \\
For any $q$ , the sum of the moduli of the two non-zero entries in
any  row of the matrix $M(x , 1)$ equals two, therefore by
Gershgorin circle theorem, all eigenvalues are in the disk
$|E| \leq 2$. 
We find that for any  $N$ the eigenvalues of 
the "q-roots of unity" model are inside
the regular polygon with $2 \, q$ sides. The derivation of this very
unusual boundary, 
together with the symmetry properties of the eigenvalue distribution 
are of intrinsic interest and are given in Sect.2. 
 In the
limit $q \to \infty $ , which corresponds to assuming the random 
variables  $x_i$  to be random phases, the support is the disk
centered in the origin, with radius equal two and the
density only depends on $|E|$.
This case is the "clock model"  of ref.\cite{fz1}.
Our second result is
the evaluation of the moments $<$ tr $M^n \,>$ by mapping  it  into
the problem of
 counting the returns to the origin of a restricted class of one-dimensional
random walks, in Sect.3 \cite{prepr}. The number of  random walks in one
 dimension that originate at a given point, which we  call
the origin, and after $2n$ random steps of unit length to the right or to the left,
 return to  the origin (not necessarily for the first time) is  $( 2 n )!/(n!)^2 .$  
However, if we select among them
the walks where each site different from the origin is visited an even number of times,
the walks have to consist of a number of steps multiple of $4$ and their number is
smaller. Let us call such random walks {\bf the even-visiting
walks}.
In the limit $N \to \infty$ , the number of even-visiting walks  provides the 
moments $<$ tr $M^n \,>$ of the "sign model".  In the same way, the
number of returns to the origin of one dimensional walks where each
site of the walk is visited a number of times multiple of  $q$
provides the moments $<$ tr $M^n \,>$ of the "q-roots of unity" model.\\

As  we mentioned , we do not know physics models related to
the "sign model". It may then be proper to spend a few words to justify
the present investigation. The evaluation of moments $<$ tr $M^n \,>$,
where the tridiagonal matrix $M(x, 1)$ is given in eq.(\ref{a.1}),
is conveniently mapped into a random walk problem in one dimension,
independently of the probability densities of the random variables
${x_i}$. Furthermore, provided the ${x_i}$ are real  independent variables
and the probability densities are even functions, only  the even-visiting
walks are relevant. However, only for the "sign model"  the evaluation
of the  moments $<$ tr $M^n \,>$ corresponds to counting
the even-visiting walks, whereas if a different even probability density
is considered, to each even-visiting walk one has to assign a weigth.\\

Our main result , for the "sign model" is
\begin{eqnarray}
 \lim_{N \to \infty} \frac{1}{N} <{\rm tr} \, M^{4 k} >=
\sum_{t=1}^k  \sum_{ \{n_i\} } S_{[n_1, n_2, \cdots , n_t ]}
\label{i.2}
\end{eqnarray}
where the sum $ \sum_{ \{n_i\} }$ is over the $t$ positive integers $n_i$ with
the restriction $n_1+n_2+\cdots +n_t=k$ and $ S_{[n_1, n_2, \cdots , n_t ]}$
is the number of the relevant walks with "width" $w=t$ :
\begin{eqnarray}
S_{[n_1, n_2, \cdots , n_t ]}=
 \frac{ 2k}{n_1}  \prod_{i=1}^{t-1}
 \left( 2 n_{i+1}+2 n_{i}-1  \atop 2 n_{i+1} \right)
\label{i.3}
\end{eqnarray}
A very similar result, for arbitrary value of $q$, for the
"q-roots of unity" model, is  obtained in Sect.3.\\

As this work was completed, a new work appeared \cite{bd}
where the spectral density of the "clock model" was obtained.
Since our work does not provide quantitative information on
spectral densities, it has little overlap with it.
However several very interesting assertions , like the analysis
of the two non-linear maps $T_+$ , $T_-$, and the existence of
bound states  , discussed
for  the "clock model", are equally valid for the "sign model".

\section{The q-roots of unity  model}

We consider the ensemble ${\cal E}_N(q,1)$ of tridiagonal random matrices  
of size $N\times N$, 
\begin{eqnarray}
   M(x,1)
  \,=\,  \left(  \begin{array}{cccccccc}
   0   & x_1 & 0   & 0   & 0 &   \cdots & 0 & 0 \\
   1 & 0   & x_2 & 0   & 0 &  \cdots & 0 & 0\\
   0   &  1 & 0   & x_3 & 0 &   \cdots & 0 & 0 \\
   0   & 0   &  1 & 0   & x_4 &   \cdots & 0 & 0 \\
  \cdots &  \cdots &  \cdots &  \cdots &  \cdots &  \cdots & 
                                                            \cdots &  \cdots \\
   0 & 0 & 0 & 0 & 0 &   \cdots & 0 & x_{N-1} \\   
   0 & 0 & 0 & 0 & 0 &    \cdots  & 1  & 0\\   
\end{array}   \right)  
\label{a.1}
\end{eqnarray}
where the random entries $x_i$, $i=1, 2,.., N-1$, are  $q-$roots of unity:
$x_i^q=1$. More precisely, they are independent and identically
distributed random variables, with uniform probability distribution over the 
set of the $q-$roots of unity:
\begin{eqnarray}
  P(x)=\frac{1}{q} \sum_{j=0}^{q-1}  \delta(x-w_q^j), \quad \quad 
w_q\equiv e^{i{{2\pi}\over q}}
\label{a.2}
\end{eqnarray}
The ensemble consists of $q^{N-1}$ different matrices.\\
For later convenience, we introduce the notation $M(x,y)$ to denote a general
tridiagonal matrix $M$ of order $N$ with the upper and lower diagonals
specified by vectors $x=\{x_1,..,x_{N-1}\}$ and $y=\{y_1,..,y_{N-1}\} $.\\

In this section we are interested in the investigation of the
symmetries and boundaries of the eigenvalue distribution of the
ensemble, in the large $N$ limit. 
We consider , for finite N, the set  $\sigma (q,N)$ in the complex plane of 
all the eigenvalues of matrices belonging to the ensemble and show that it 
has nice symmetry properties.\\

{\underbar {Proposition 1}}: The set $\sigma (q,N)$ is invariant under the
transformations\\
\centerline { i) $E\to -E$, \quad ii) $E\to E^*$, \quad iii) 
$E\to Ee^{i{\pi\over q}}$ }
Proof:  the characteristic polynomial of a generic tridiagonal matrix $M(x,y)$
only depends on the products $x_iy_i$, $i=1,2,..,N-1$. Therefore, for a given  
matrix $M(x,1)$ in the ensemble we introduce the two matrices $M(z,z)$ and 
$M(-z,-z)$ where $z_i^2=x_i$. Clearly they have opposite eigenvalues, meaning
that both $E$ and $-E$ are eigenvalues of $M(x,1)$. \parno
The ensemble ${\cal E}_N(q,1)$ is closed under complex conjugation, and
if $\{ E_i\}$ are the eigenvalues of $M(x,1)$, $\{E_i^*\}$ are the eigenvalues
of $M(x^*,1)$.\parno
The set of matrices $M(z,z)$, where each $z_i$ is randomly chosen in the set 
of the $2q-$roots of unity, is invariant under multiplication of a matrix by 
the scalar $e^{i\pi/ q}$. Therefore, if $\{E_i\}$ are the eigenvalues of 
$M(x,1)$, $\{ E_i e^{i\pi/q}\}$ are the eigenvalues of $M(xw_q,1)$.\parno
We conclude that the set $\sigma (q,N)$ has the above stated symmetries
(note that invariance i follows from iii). $\bullet $\\

{\underbar {Proposition 2}}: The set $\sigma (q,N)$ is contained in the
regular $2q-$polygon in the complex $E-$plane, with corners 
$E_k=2 e^{ik\pi /q} $, $k=0,1,..,2q-1$.\parno
Proof. Let $E$ be an eigenvalue of a given matrix $M(x,1)$ in the ensemble;
because of the symmetry, we assume that $0\le {\rm {arg}}E \le \pi /q$. 
The eigenvalue equation for $M(x,1)$ is $u_{i-1}+x_i u_{i+1}=Eu_i $, 
with $u_0=u_{N+1}=0$. Let $I_k$ be the set of integers $j=1,..,N-1$ such that 
$x_j=w_q^k$. After multiplication by $u_i^*$ of the eigenvalue equation and
summation over $i$:
\begin{eqnarray}
  E&=&\sum_{i=1}^N u_i^* u_{i-1} + \sum_{i=1}^N x_i u_i^*u_{i+1}=\nonumber \\
     &=&\sum_{i=1}^N u_{i+1}^* u_i + \sum_{k=1}^q w_q^k 
                            \sum_{i\in I_k} u_i^*u_{i+1}= \nonumber \\
     &=&\sum_{k=1}^q \sum_{i\in I_k} u_{i+1}^*u_i+ w_q^k u_i^*u_{i+1}=
\nonumber \\
     &=&\sum_{k=1}^q e^{ik{\pi\over q}}
             \sum_{i\in I_k} \left (u_i^*u_{i-1}e^{-ik{\pi\over q}}+ c.c. 
\right ). \nonumber \\ 
\nonumber
\end{eqnarray}
We finally take the real and imaginary parts of $E$ and build the 
majorization:
\begin{eqnarray}
  && {\rm Re}E \cos (\frac {\pi}{2q}) + {\rm Im}E \sin (\frac {\pi}{2q}) =
\sum_{k=1}^q \cos (k\frac {\pi}{q} - \frac {\pi}{2q})
\sum_{i\in I_k} \left (u_i^*u_{i-1}e^{-ik{\pi\over q}}+ c.c. \right ) 
\nonumber \\
&& \le  2 \sum_{k=1}^q \vert \cos (k\frac {\pi}{q} - \frac {\pi}{2q}) \vert
\sum_{i\in I_k} |u_i^*u_{i-1}| \le  2 \cos ( \frac {\pi}{2q}) 
\sum_{k=1}^q \sum_{i\in I_k} 
 |u_i^*u_{i-1}| \nonumber \\
&& \le 2 \cos ( \frac {\pi}{2q}) \nonumber \\ \nonumber 
\end{eqnarray}
Therefore: ${\rm Im}E\le {\rm cotg} (\frac {\pi}{2q}) (2- {\rm Re}E )$, which
ends the proof. $\bullet $ \\

From the discussion of proposition 1, we conclude that the distribution of
the eigenvalues $\rho (E,E^*)$ for the ensemble ${\cal E}_N(q,1)$ 
has the listed symmetries, and therefore it is a symmetric function of the 
variables $E^{2q}$ and $(E^*)^{2q}$. This puts contraints on the moments.\\

{\underbar {Proposition 3}}:  $\langle E^m E^{*n}\rangle \neq 0 $ only if
$m+n$ is even and $|m-n|=2rq$.\\
Proof: Since the expectation value is symmetric in $m$ and $n$, we put 
$m\ge n$. Due to the symmetry $E\to -E$ of the density, non-vanishing of the
expectation value requires $m+n=2s$, so that $m-n=2\ell$. By the $2q-$fold 
symmetry of the density and the support $\sigma $, we may evaluate the 
expectation by integrating over $\omega $, the angular sector 
$0\le {\rm arg}E\le \pi/q$ of $\sigma $:
\begin{eqnarray}
\langle |E|^{2s}E^\ell (E^*)^{-\ell}\rangle = \int_\omega 
d^2E \, \rho (E,E^*) |E|^{2s}
 \sum_{k=0}^{2q-1}(Ee^{ik\pi/q})^\ell (E^*e^{-ik\pi /q})^{-\ell}
\nonumber
\end{eqnarray}
We get a non-zero result for the sum only if $\ell =rq$. $\bullet $\\

In the limit $q\to\infty $, where each number $x_i$ is an arbitrary number on
the unit circle, the set $\sigma (\infty, N)$ is invariant under complex
rotations, and the eigenvalue density is a function of $|E|$.\\

\section{The moments and counting the returns of the relevant walks.}

In this section we study a set of moments of the probability density
$\rho (E,E^*)$ of the "q-roots of unity" model, which have the
interesting interpretation as counting numbers of the random walks on
the line that return to the origin and visit each intermediate site
a number of times which is multiple of $2q$.
These moments have generating function
\begin{eqnarray}
G(z)=  \frac{1}{N}  <{\rm tr} \, \frac{1}{z-M}>
\label{c.1}
\end{eqnarray}
where the expectation value is evaluated on the independent identically
distributed random variables $x_i$ :
\begin{eqnarray}
< f(M)>= \int  \prod_{i=1}^N P(x_i) \, dx_i \, f \left(M(x,1) \right)
\label{c.2}
\end{eqnarray}
The formal perturbative expansion of $G(z)$ is
\begin{eqnarray}
G(z)= \frac{1}{N}  \sum_{k=0}^{\infty}  \frac{ <{\rm tr}\, M^k >}{z^{k+1}}
\label{c.3}
\end{eqnarray}

Since we are interested in the limit $N \to \infty$, any term
on the diagonal  $< [M^k]_{r r} >$ has the same value, and the trace
merely cancels the $1/N$ factor.

It is useful to consider the one to one map between the non vanishing terms 
contributing to $ [M^k]_{a b}$ and the random walks in one dimension which
 originate at site $a$ to arrive at site $b$ after $k$ steps.

\begin{figure}[t]
\begin{center}
\mbox{\epsfig{file=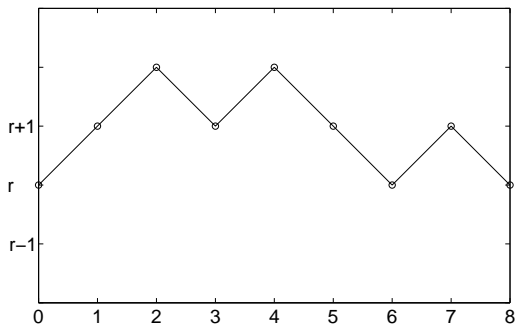,height=5cm}}
\end{center}
\caption{One of the even-visiting random walks, returning to site $r$ after $8$ 
steps, with width $w=2$. }
\end{figure}

\begin{figure}[t]
\begin{center}
\mbox{\epsfig{file=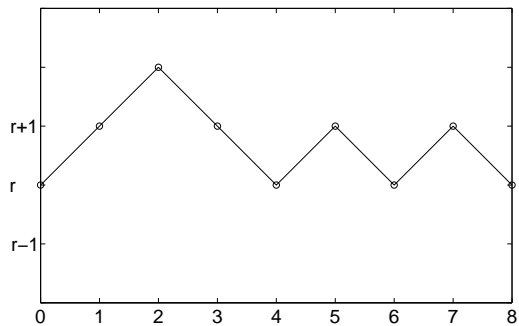,height=5cm}}
\end{center}
\caption{One of the  random walks not belonging to the class of even-visiting
 random walks. }
\end{figure}

 Let us consider the term $k=8$
\begin{eqnarray}
\sum_{a,b,c,d,e,f,g} < M_{ra} M_{ab} M_{bc} M_{cd} M_{de} M_{ef} M_{fg} M_{gr}>
\label{c.4}
\end{eqnarray}
By recalling that the non zero matrix elements are
 $ M_{i j}=1 $ if $j=i-1 ,$  and  $ M_{i j}=x_i $ if $j=i+1$ , each term in the sum (\ref {c.4})
corresponds to a walk of 8 steps, originating and ending at site $r$, with
4 steps up and 4 steps down. For instance, 
the sequence  \\
$M_{r,r+1}
M_{r+1,r+2} M_{r+2,r+1} M_{r+1,r+2} M_{r+2,r+1} M_{r+1,r} M_{r,r+1} M_{r+1,r}$ =
$x_r \cdot x_{r+1} \cdot 1 \cdot x_{r+1} \cdot 1 \cdot 1 \cdot x_r \cdot 1 = 
1$ is shown in Fig.1,
while the
sequence  $x_r \cdot x_{r+1} \cdot 1  \cdot 1 \cdot x_r  \cdot 1 \cdot x_r
 \cdot 1 $ = $x_r \cdot x_{r+1}$   is shown in Fig.2 .
Each walk corresponding to a product of
random variables $\prod_j (x_j)^{n_j}$ where all the powers ${n_j}$ are 
multiple of $q$  yields a contribution $+1$, while the walks where
at least one power ${n_j}$ is not a multiple of $q$ are averaged to zero ,
therefore being a class of irrelevant walks.
 Then $\frac{1}{N}
<{\rm tr} M^k> $ is equal to the number of relevant walks of $k$ steps, that
is the walks from a fixed site
$r$ to the same site $r$ , such that each intermediate site is visited a
multiple of $q$ times.

\begin{figure}[t]
\begin{center}
\mbox{\epsfig{file=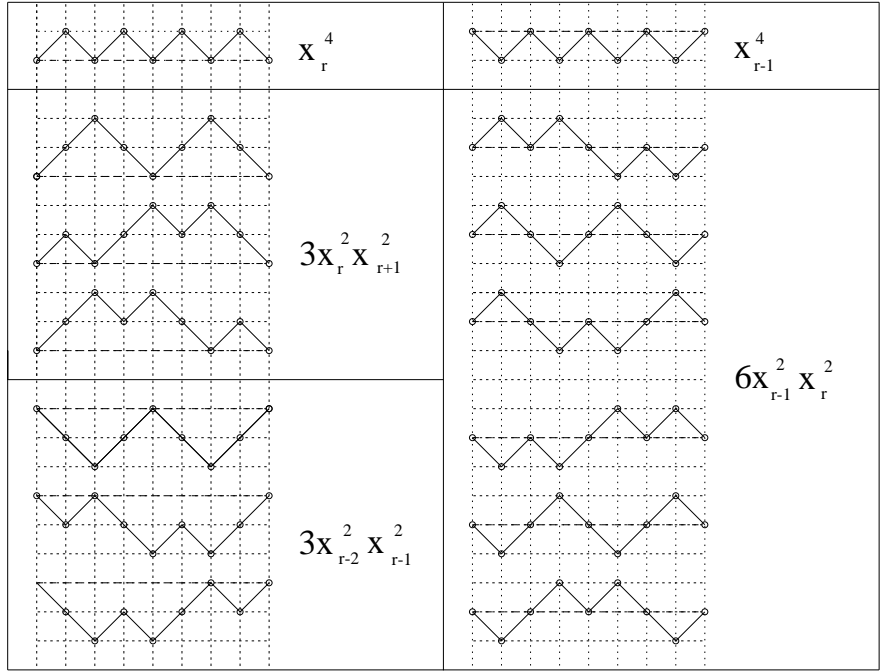,height=10cm}}
\end{center}
\caption{Here are shown the 14  paths of 8 steps, relevant for the "sign model" }
\end{figure}

 Because the number of steps up (each associated to a random variable $x_i$)
is equal to number of steps down (each associated to a factor one) , the total
number of steps $k$ of the relevant walks is a multiple of $2 q$
 and we rewrite eq.(\ref{c.3}) as
\begin{eqnarray}
G(z)= \sum_{k=0}^{\infty} \frac{c_k}{z^{2 q  k +1}} \quad ; \quad
c_k= \lim_{N \to \infty} \frac{1}{N} <{\rm tr} \, M^{2 q k} >
\label{c.5}
\end{eqnarray}
This power series is absolutely convergent for $|z|>2$. In the case of $q= \infty$
we obtain
\begin{eqnarray}
G(z)= \frac{1}{z} \qquad , \qquad {\rm for} \quad |z|>2
\label{c.6}
\end{eqnarray}
Next we present the combinatorial evaluation of the coefficients $c_k$
for the case $q=2$ , that is the number of returns to the origin
for even visiting walks. The case of general integer values of $q$ requires only
trivial generalization, presented in the following paragraph.\\

{\underbar {Counting the returns to the origin of even visiting walks }}.
We shall indicate with
 $N( 2n_{-s}, 2 n_{-s+1} , \cdots , 2 n_{-1}\, ;\, 2n_0, 2 n_{1} , \dots , 2 n_{j})$
the  number of even visiting walks from site $r$ to site $r$ corresponding
to the product  \\
$(x_{r-s})^{2n_{-s} } \cdots (x_{r-1})^{2 n_{-1}}
(x_r)^{2n_0} (x_{r+1})^{2 n_{1}}\cdots (x_{r+j})^{2 n_{j} }$. 
All the integers $n_{-s},  \dots n_j$ are positive ; the semicolon separates
the exponents $n_i$
 associated to sites  $i<r$ from those  associated to sites 
$i \geq r$. We omit strings of zeros external
to the string of positive integers, just keeping one zero for
walks visiting only sites $s \geq r$ or only sites $s \leq r$ . In the first
case we write the multiplicity as $N(0 \, ; \, 2n_0,  \cdots , 2n_j)$ ,
in the latter case we write $N(2 n_{-s}, \cdots , 2n_{-1} \, ; \,0)$.
It is easiest to
 begin with the evaluation of $N(0 \, ; \, 2n_0,  \cdots , 2n_j)$ .
The length of the walk is $4 (n_0+n_{1}+\cdots +n_{j})$. The "maximum
site" visited is the site $r+j+1$ , visited $2n_{j}$ times ; the "minimum site"
 visited is the site $r$ ,  visited $2n_0 +1$ times. \\

The number
$N(0 \, ; \, 2n_0, 2 n_{1} , \cdots , 2 n_{j} , 2 n_{j+1} )$ is related to 
 $N(0 \, ; \, 2n_0, 2 n_{1} , \cdots , 2 n_{j})$ in the  following way : new walks of
length two corresponding to $(x_{r+j+1} \cdot 1)$ may be inserted in each of the
maxima of the previous walk. Since $2 n_{j+1}$ identical objects are placed in
$2 n_{j}$ places in $ \left( 2 n_{j+1}+2 n_{j}-1 \atop 2 n_{j+1} \right)$
ways, we obtain
\begin{eqnarray}
&& N(0 \, ; \, 2n_0, 2 n_{1} , \cdots , 2 n_{j} , 2 n_{j+1} )=
\nonumber \\
&=&
 \left( 2 n_{j+1}+2 n_{j}-1 \atop 2 n_{j+1} \right)
N(0 \, ; \, 2n_0, 2 n_{1} , \cdots , 2 n_{j})
\label{b.1}
\end{eqnarray}

\begin{figure}[t]
\begin{center}
\mbox{\epsfig{file=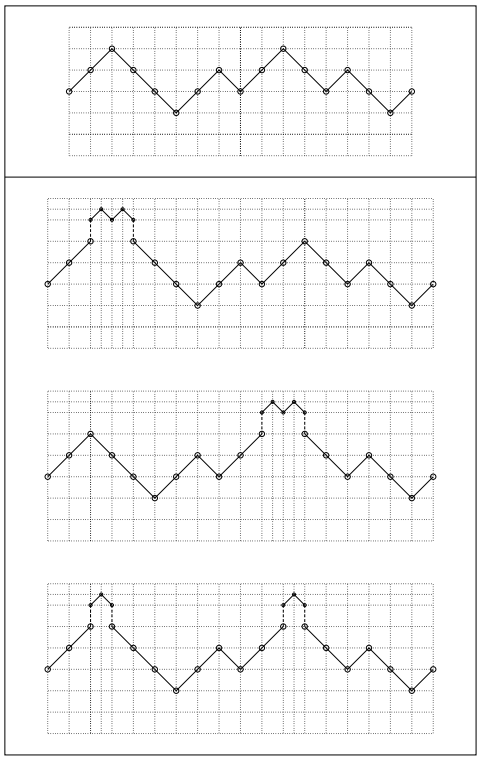,height=10cm}}
\end{center}
\caption{The figure illustrates the insertion of a 4-steps part over anyone of
the tops of a previous relevant walk of width $w$,
 thus obtaining new relevant walks,
of width $w+1$ , according to eq.(\ref {b.1}) }
\end{figure}

By iterating eq.(\ref{b.1}) with the initial condition $N(0 \, ; \, 2 n_0)=1$ one
obtains
\begin{eqnarray}
N(0 \, ; \, 2n_0, 2 n_{1} , \cdots , 2 n_{j}) = \prod_{i=0}^{j-1}
 \left( 2 n_{i+1}+2 n_{i}-1  \atop 2 n_{i+1} \right)
\label{b.2}
\end{eqnarray}

We proceed to evaluate the multiplicity of the relevant random walks
that visit sites $r , \, r+1, \cdots ,r+j$ as often as before, and in addition
visit $2 n_{-1}$ times the site $r-1$ . Each walk of this class
may be obtained by inserting 
 $2 n_{-1}$ 
walks of length two $(1 \cdot x_{r-1})$ in each of the
 $ 2 n_0 +1$  minima of
the walk of the previous class.
Therefore
\begin{eqnarray}
&& N(2 n_{-1} \, ; \, 2 n_0, 2 n_{1} , \cdots , 2 n_{j})=
\nonumber \\
&=&  \left( 2 n_{-1}+2 n_0 \atop 2 n_{-1} \right)
N(0 \, ; \, 2n_0, 2 n_{1} , \dots , 2 n_{j} )
\label{b.3}
\end{eqnarray}
The procedure may be repeated to include the relevant walks which
visit sites $r-2, \, r-3,\cdots , r-s$. We obtain
\begin{eqnarray}
&& N(2 n_{-s} , 2 n_{-s+1}, \cdots, 2 n_{-1} \, ; \, 2 n_0, 2 n_{1} , \cdots , 2 n_{j})=
\nonumber \\ &=&  \left[
\prod_{p=0}^{s-2} \left(  2 n_{-s+p}+2 n_{-s+p+1} -1 \atop 2 n_{-s+p} \right)
\right] \left( 2 n_{-1}+2 n_0 \atop 2 n_{-1} \right)
 \left[\prod_{i=0}^{j-1}
 \left( 2 n_{i+1}+2 n_{i}-1  \atop 2 n_{i+1} \right) \right]
\nonumber \\
\label{b.4}
\end{eqnarray}

The coefficient $c_k$  is the number of the even visiting walks of $4 k$
steps and it is the sum of the multiplicities
$N(2 n_{-s} , 2 n_{-s+1}, \cdots, 2 n_{-1}\, ; \, 2 n_0, 2 n_{1} , \dots , 2 n_{j})$
given above, where  $ k= n_{-s} +  n_{-s+1}+ \cdots + n_{j} $.\\

The evaluation may be someway simplified, by considering 
walks of fixed width, that is the difference between the "maximum site"
visited and the "minimum site" visited .
We consider the set of ordered partitions
of $k$ into positive integers $ [n_1, n_2, \cdots , n_t]$ where $k=\sum n_p$.
 Each ordered sequence , 
 $ [n_1, n_2, \cdots , n_t]$  , corresponds to $t+1$ classes
of even visiting walks, which are associated to the products
\begin{eqnarray}
&& (x_{r-t})^{2 n_1} (x_{r-t+1})^{2 n_2} \cdots  (x_{r-1})^{2 n_t} \quad ; 
\nonumber \\
&& (x_{r-t+1})^{2 n_1} (x_{r-t+2})^{2 n_2} \cdots  (x_{r})^{2 n_t} \quad ; 
\nonumber \\
&&\qquad
 \cdots \qquad \cdots \qquad  \cdots \qquad   \cdots \qquad ; \nonumber \\
&& (x_{r})^{2 n_1} (x_{r+1})^{2 n_2} \cdots  (x_{r+t-1})^{2 n_t} 
\label{b.5}
\end{eqnarray}
All walks in eq.(\ref{b.5}) have the same width $w=t$. Their multiplicities,
given in eq.(\ref{b.4}) are simply related and their sum is
\begin{eqnarray}
S_{ [n_1, n_2, \cdots , n_t ] }= \frac{ 2k}{n_1}  \prod_{i=1}^{t-1}
 \left( 2 n_{i+1}+2 n_{i}-1  \atop 2 n_{i+1} \right) \quad ; \quad
S_{ [n_1 ] }= 2
\label{b.6}
\end{eqnarray}

Next we sum over the ordered partitions
$ [n_1, n_2, \cdots , n_t]$ of $k$ into $t$ parts and
 finally over the different widths,
from $1$ to $k$ 
\begin{eqnarray}
c_k=\sum_{t=1}^k  \sum_{ \{n_i \}} S_{[n_1, n_2, \cdots , n_t ]} 
\label{b.7}
\end{eqnarray}
where the sum $ \sum_{ \{n_i \}}$ is over the $t$ positive integers $n_i$
with the restriction $n_1+n_2+\cdots+n_t=k$.
The evaluation of eq.(\ref{b.7}) may be automated and we find
 the first coefficients $c_k$  :
\begin{eqnarray}
\begin{tabular}{||r | r ||} \hline
$c_0$ & 1       \\  \hline
$ c_1$& 2       \\  \hline
$ c_2 $ & 14      \\  \hline
$c_3$ & 116    \\  \hline
$c_4$ & 1 \,110  \\  \hline
$c_5$ & 11 \, 372 \\  \hline
$c_6$ & 123 \, 020 \\  \hline
$c_7$ & 1 \, 384 \, 168 \\  \hline
$c_8$ & 16 \, 058 \, 982 \\  \hline
$c_9$ & 190 \, 948 \, 796 \\  \hline
$c_{10}$ & 2 \,317 \, 085 \, 924 \\  \hline
$c_{11}$ & 28 \, 602 \, 719 \, 576 \\  \hline
$c_{12}$ & 358 \, 298 \, 116 \, 092 \\  \hline
$c_{13}$ & 4 \, 545 \, 807 \, 497 \, 272 \\  \hline
$c_{14}$ & 58 \, 321 \, 701 \, 832 \, 408 \\  \hline
$c_{15}$ & 755 \, 700 \, 271 \, 652 \, 816  \\  \hline
$c_{16}$ & 9 \, 878 \, 971 \, 460 \, 641 \, 414 \\  \hline
\end{tabular}
\label{b.8}
\end{eqnarray}

\begin{figure}[t]
\begin{center}
\mbox{\epsfig{file=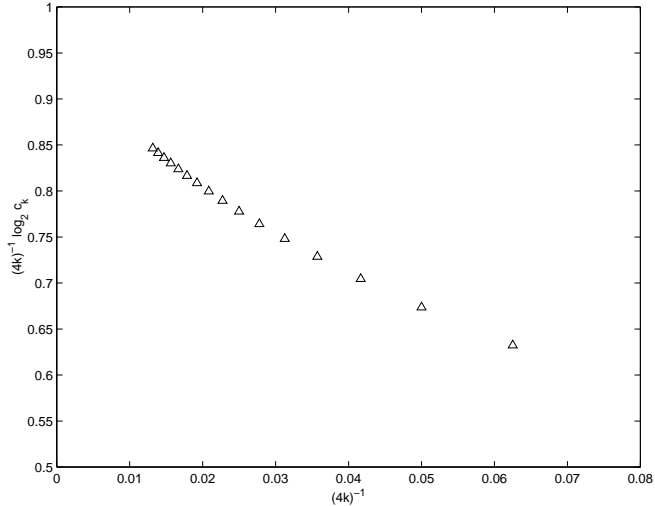,height=6.8cm}}
\end{center}
\caption{The triangles are the values of $ (1/ 4k) \, \log_2 \, c_k$ versus
$ (1/ 4k) $, for $k=4$ to $19$. The convergence for large values of $k$
is consistent with $ c_k \sim a^k$ , with $a \leq 16 $}
\end{figure}

The ratios $c_{n}/c_{n-1}$ rise monotonically with a rate slower at higher
values of $n$. We know from the previous section that the
eigenvalues of the random matrix $M$   are inside the square 
 with vertices in the four points
$\pm 2 $ , $  \pm 2i $ , it then follows that $c_n \sim a^n$ , with   $a \leq 16 $ .
The plot of the coefficients $c_n$ for $n=4$ to $19$ in Fig.5 is
consistent with the expected asymptotics.  $\bullet $\\

{\underbar {Counting the returns to the origin of  walks for generic q }}.
The combinatorial evaluation given above holds with trivial modifications for the
"q-roots of unity" model, described in eqs.(\ref{a.1}) , (\ref{a.2}).
Again tr $M^k $ corresponds to one-dimensional walks returning to the original
site after $k$ steps, each walk being the product of $k/2$ random variables $x_i$.
The average of each product, with the factorized probability distribution
(\ref{a.2}) vanishes unless each random variable occurs with a power multiple of
$q$ , in this case the product has average one. Therefore $k/2$ is  a multiple
of $q$ and the class of relevant walks is such that each site (apart the origin of
the walk) is visited a number of times multiple of $q$. 
Let $N(0 \, ; \, qn_0, qn_{1},..,q n_{j})$ 
be the number of the relevant walks
corresponding to the multiplicity of the product $(x_r)^{q n_0} 
(x_{r+1})^{q n_{1}}
\dots (x_{r+j})^{q n_{j} }$. The insertion above the top sites or below the bottom
ones proceeds as before leading to a trivial generalization for the
number of relevant walks $S_{ [n_1, n_2, \cdots , n_t ] }$ having
fixed $ \rm{width }\; t$
\begin{eqnarray}
S_{ [n_1, n_2, \cdots , n_t ] }= \frac{ 2k}{n_1}  \prod_{i=1}^{t-1}
 \left( q n_{i+1}+q n_{i}-1  \atop q n_{i+1} \right)
\label{b.9}
\end{eqnarray}

\begin{eqnarray}
c_k=\sum_{t=1}^k  \sum_{ \{n_i \} } S_{[n_1, n_2, \cdots , n_t ]} 
\label{b.10}
\end{eqnarray}
where the sum $ \sum_{ \{n_i\} }$ is over the $t$ positive integers $n_i$ with
the restriction $n_1+n_2+\cdots +n_t=k$ . $\bullet $ \\

Let us close with few words for the special value $q=1$. In this case the
matrix $M$ in eq.(\ref{a.1}) is no longer random, it is the real
symmetric matrix with all $x_i=1$. Its spectral distribution is well known and the
analysis in random walks is not necessary. Still one may find that
$c_k=$ tr $M^k(1,1)=\left( 2k \atop k \right)$ is the number of one-dimensional
random walks returning to the origin after $2k $ steps. We checked , for several
values of $k$, that eqs.(\ref{b.9}) , (\ref{b.10}) reproduce the correct value for
 $q=1$.

\vskip 1cm
\centerline{\bf Acknowledgment}

We wish to thank prof.A.Zee who introduced us to the "sign model"
and encouraged this study.

\end{document}